\documentclass[12pt]{article}
\usepackage{epsfig}
\usepackage{color}
\usepackage{amssymb,amsmath}
\usepackage{graphicx}
\usepackage{epsfig}

\newcommand{\bea}{\begin{eqnarray}}
\newcommand{\eea}{\end{eqnarray}}

\newcommand{\mpt}{p\hspace{-0.45em}/ }

 \makeatletter
\def\alt{\mathrel{\mathpalette\gl@align<}}
\def\agt{\mathrel{\mathpalette\gl@align>}}
\def\gl@align#1#2{\lower.6ex\vbox{\baselineskip\z@skip\lineskip\z@
\ialign{$\m@th#1\hfil##\hfil$\crcr#2\crcr\sim\crcr}}} \makeatother

\begin{document}
\begin{flushright}
KEK-TH-1645
\end{flushright}
\vspace*{1.0cm}

\begin{center}
\baselineskip 20pt 
{\Large\bf 
An equal-velocity scenario for hiding dark matter at the LHC
}
\vspace{1cm}

{\large 
Kaoru Hagiwara$^{a, b}$, Toshifumi Yamada$^a$
} \vspace{.5cm}

{\baselineskip 20pt \it

$^a$ KEK Theory Center, \\
1-1 Oho, Tsukuba, Ibaraki 305-0801, Japan \\

$^b$ SOKENDAI, \\
1-1 Oho, Tsukuba, Ibaraki 305-0801, Japan \\

}

\vspace{.5cm}

\vspace{1.5cm} {\bf Abstract} \end{center}
We consider a light supersymmetric top quark (\textit{stop}) scenario,
 where a \textit{stop} and a neutralino are the lightest supersymmetric particles
 and the \textit{stop} mass is close to the sum of the neutralino mass and 
 the top quark mass.
In this scenario, the top quark and the neutralino coming from the decay of a \textit{stop}
 have almost equal velocity vectors,
 and hence the missing transverse momentum in \textit{stop} pair production events at colliders
 is suppressed, which makes the \textit{stop} search challenging,
 as in the degenerate scenario where the \textit{stop} mass and the neutralino mass
 are nearby.
In this paper, we propose a novel analysis technique
 aiming at discovering the \textit{stop} pair production signal in such an `equal-velocity' scenario.
The key is to look for events where the missing transverse momentum vector is
 proportional to the transverse momentum vector of the top quark pair
 with specific values for the proportionality coefficient,
 as this coefficient equals to the ratio of the neutralino mass over the top quark mass
 in the equal-velocity limit.
We examine the possibility that the \textit{stop} signal can be identified
 at the $\sqrt{s}=14$~TeV LHC by our analysis method.

\thispagestyle{empty}

\newpage

\setcounter{footnote}{0}
\baselineskip 18pt
%

\ \ \ Although the gauge hierarchy problem suggests that
 the standard model (SM) should be extended at $O(100)$ GeV scale,
 the 8~TeV LHC has reported no signal of new physics so far.
One possible explanation to this gap is that new physics signatures are hidden behind
 SM backgrounds due to the lack or smallness of missing transverse momentum,
 which is often utilized for tagging new physics events with a dark matter candidate.
In this letter, we consider a scenario based on the R-parity conserving 
 minimal supersymmetric standard model (MSSM),
 where the lightest R-parity odd particle is a dark matter particle, and
 the next-to-lightest R-parity odd particle has the mass which approximately equals to
 the sum of the masses of its SM partner and the dark matter particle 
 (the lightest R-parity odd particle).
In MSSM, we can consider the following four cases:
\begin{align*}
m_{\tilde{t}_1} \ &\simeq \ m_t \ + \ m_{\tilde{\chi}_1^0} , & {\rm (1a)} 
\\
m_{\tilde{\chi}_2^0} \ &\simeq \ m_h \ + \ m_{\tilde{\chi}_1^0} , & {\rm (1b)}
\\
m_{\tilde{\chi}_2^0} \ &\simeq \ m_Z \ + \ m_{\tilde{\chi}_1^0} , & {\rm (1c)}
\\
m_{\tilde{\chi}_2^{\pm}} \ &\simeq \ m_W \ + \ m_{\tilde{\chi}_1^0} , & {\rm (1d)}
\end{align*}
 where $\tilde{\chi}_1^0$ denotes the lightest neutralino which is a dark matter particle.
A common feature of these cases is that the heavy SM particle $t, \, h, \, Z$ or $W$
 and the dark matter particle $\tilde{\chi}_1^0$ have similar velocity vectors once they are produced
 from the decay of the next-to-lightest R-odd particle.
At colliders, therefore, the missing transverse momentum coming from the dark matter particles
 is suppressed in events where next-to-lightest R-odd particles are pair-produced 
 and then decay.
Nevertheless, such events can be identified through
 the information that the missing transverse momentum vector due to the dark matter particles
 and the transverse momentum vector of the heavy SM particles are parallel, and
 that the ratio of their absolute values equals to the ratio of the dark matter mass
 and the SM particle mass.
This is the key observation of this letter.
Although the mass spectra like eqs.~(1a-1d) are realized only for very particular combinations of 
 the model parameters,
 we have an example in the SM
 where $D^*$ particles can be identified by requiring that a $D$ candidate and a $\pi$ have
 the same velocity vector because the mass of $D^*$ is close to the sum of the $D$ mass and the $\pi$ mass.
We hereafter call the scenario the `equal-velocity scenario'.
\setcounter{equation}{1}

In this letter, particular attention is paid to the mass spectrum of eq.~(1a).
This spectrum is motivated by the light SUSY top (\textit{stop}) scenario \cite{light stop},
 which is inspired by the fact that the top quark and its SUSY partners
 have the largest couplings with the Higgs boson
 and hence the SUSY top partners should be light in order to solve the gauge hierarchy problem.
Also, since the \textit{stop}s are colored particles,
 it is relatively easy to find the \textit{stop} signal at the LHC,
 compared to the cases with other mass spectra eqs.~(1b-1d).
Search strategies and techniques for the light \textit{stop} scenario
 have been discussed in refs.~\cite{stop search}.
In this study, we focus on a mass spectrum where the \textit{stop} mass is close to the sum of the top quark mass and the mass of
 the lightest R-odd particle, and propose a novel search method optimized for such a mass spectrum.

Before describing our new search method,
 we review the status of light \textit{stop} searches in the LHC 7~TeV and 8~TeV runs,
 concentrating on the case where a \textit{stop} and a neutralino are the only light SUSY particles
 and their mass difference is nearly equal to or larger than the top quark mass.
Searches for such a mass spectrum have been done
 by observing events with hadronic \cite{lhchad}, semi-leptonic \cite{lhcsemi} or di-leptonic decays of a top quark pair \cite{lhcdi}
 associated with large missing transverse momentum.
With about 20~fb$^{-1}$ of data at the 8~TeV LHC, the parameter region with 200~GeV $< m_{\tilde{t}_1} < 300$~GeV and $m_{\tilde{\chi}_1^0} \lesssim (m_{\tilde{t}_1} - m_t - 15$~GeV)
 and the region with 300~GeV $<m_{\tilde{t}_1} \lesssim 600$~GeV and $m_{\tilde{\chi}_1^0}$ much smaller than $(m_{\tilde{t}_1} - m_t - 15$~GeV)
 are excluded.
Additionally, a search using the spin correlation of a top quark pair has been conducted \cite{lhcspin},
 and the region with $m_t<m_{\tilde{t}_1} \lesssim 195$~GeV and $m_{\tilde{\chi}_1^0}< (m_{\tilde{t}_1} - m_t)$ is excluded.
\\

In the equal-velocity limit where $m_{\tilde{t}_1}=m_t+m_{\tilde{\chi}^0_1}$ holds,
 the top quark and the neutralino originating from the decay of a \textit{stop}
 have the same velocity vector.
Hence, in a \textit{stop} pair production event at hadron colliders
\begin{eqnarray}
p p &\rightarrow& \tilde{t}_1 \bar{\tilde{t}}_1 \ \rightarrow \ t \tilde{\chi}^0_1 \bar{t} \tilde{\chi}^0_1 \ ,
\end{eqnarray}
 the top quarks and the neutralinos have the following three-momenta:
\begin{align}
\vec{p}_t \ &= \ m_t \gamma_1 \vec{\beta}_1 \ , \ \ \ \vec{p}_{\bar{t}} \ = \ m_t \gamma_2 \vec{\beta}_2 \ ,
\ \ \ \vec{p}_{(\tilde{\chi}^0_1)^a} \ = \ m_{\tilde{\chi}^0_1} \gamma_1 \vec{\beta_1} \ , \ \ \ 
\vec{p}_{(\tilde{\chi}^0_1)^b} \ = \ m_{\tilde{\chi}^0_1} \gamma_2 \vec{\beta}_2 \ ,
\end{align}
 where the indices $a$ and $b$ label the two neutralinos,
 $\vec{\beta}_1$ and $\vec{\beta}_2$ respectively denote the velocity vectors
 of $\tilde{t}_1$ and $\bar{\tilde{t}}_1$ in the laboratory frame,
 and $\gamma_1$ and $\gamma_2$ denote their corresponding boost factors.
It follows that the missing transverse momentum vector coming from the neutralinos, $\vec{\mpt}_T$,
 and the transverse momentum vector of the top quark pair, $\vec{p}_{t\bar{t} \, T}$, 
 satisfy the following relation in the equal-velocity limit:
\begin{align}
\vec{\mpt}_T \ &= \ (\vec{p}_{(\tilde{\chi}^0_1)^a}+\vec{p}_{(\tilde{\chi}^0_1)^b})_T \ = \ 
\frac{m_{\tilde{\chi}^0_1}}{m_t} (\vec{p}_t+\vec{p}_{\bar{t}})_T \ = \
\frac{m_{\tilde{\chi}^0_1}}{m_t} \vec{p}_{t\bar{t} \, T} \ .
\label{momenta relation}
\end{align}
To utilize the relation eq.~(\ref{momentum relation}) to distinguish \textit{stop} events from SM backgrounds in the equal-velocity scenario, 
 we plot each event on the 2-dimensional plane spanned by
 the azimuthal angle between the missing transverse momentum vector and the reconstructed top-pair transverse momentum vector,
 $\Delta \phi( \vec{\mpt}_T, \, \vec{p}_{t\bar{t} \, T} )$,
 and the ratio of their absolute values, $\mpt_T/p_{t\bar{t} \, T}$.
As the mass spectrum approaches to the equal-velocity limit,
 \textit{stop} pair production events gather around the point of 
 $\Delta \phi( \vec{\mpt}_T, \, \vec{p}_{t\bar{t} \, T} )=0$ and $\mpt_T/p_{t\bar{t} \, T}=m_{\tilde{\chi}^0_1}/m_t$,
 whereas SM background events that mainly come from top quark pair production events do not show such a tendency.

When the mass spectrum deviates from the equal-velocity limit, with
 the value of $\Delta m \equiv m_{\tilde{t}_1} - m_t - m_{\tilde{\chi}^0_1}$ being as large as $O$(1)~GeV,
 the top quark and the neutralino from one \textit{stop} decay have slightly different velocity vectors.
Even in the equal-velocity limit, the sizable top quark decay width of about 2~GeV causes
 separation of the velocity vectors of the reconstructed top quark and the neutralino
 because the \textit{stop} decay via off-shell top quark does not satisfy the kinematic condition that leads to eq.~(\ref{momenta relation}).
These departures from the equal-velocity limit can be parametrized by the 3-momentum of the reconstructed top quark
 in the rest frame of the mother \textit{stop}, which we denote by $\vec{P}$.
The absolute value of $\vec{P}$, $P$, is the solution to eq.~(\ref{P-sol}),
\begin{align}
m_{\tilde{t}_1} \ = \ m_t + m_{\tilde{\chi}^0_1} + \Delta m \ &= \ 
\sqrt{P^2 + m_{\tilde{\chi}^0_1}^2} + \sqrt{P^2 + m_t^2 - x \Gamma_t m_t}
\label{P-sol}
\end{align}
 where $x$ measures the off-shellness of the top quark propagator in the \textit{stop} decay and
 follows the probability density function $f(x)$ and the range of eq.~(\ref{prob}),
\begin{align}
f(x) \, {\rm d}x \ &\propto \ P(x)^2 \, \frac{{\rm d}P(x)}{{\rm d}x} \, \frac{1}{x^2+1} \, {\rm d}x,
\nonumber \\
- \frac{2 m_t \Delta m + \Delta m^2}{m_t \Gamma_t} \, &< \, x \, < \, \frac{2 m_{\tilde{t}_1} m_t + 2 m_{\tilde{t}_1} \Delta m - 2 m_t \Delta m - \Delta m^2}{m_t \Gamma_t}
\label{prob}
\end{align}
 where $P(x)$ in turn depends on $x$ through eq.~(\ref{P-sol}).
For $\Delta m \ll m_t, m_{\tilde{\chi}^0_1}$ and $\vert x \vert \lesssim 10$, the solution to eq.~(\ref{P-sol}) is approximately given by
\begin{align}
P \ &\simeq \ \sqrt{ \frac{ m_t }{ m_{\tilde{t}_1} } ( m_{\tilde{t}_1} - m_t ) } \sqrt{ 2\Delta m - x \Gamma_t } \ ,
\end{align}
 and the probability density function for $P$ is expressed as
\begin{align}
\bar{f}(P) \, {\rm d}P \ &\propto \ P^2 \, \frac{1}{\left(P^2  - 2\Delta m \, \frac{m_t(m_{\tilde{t}_1}-m_t)}{m_{\tilde{t}_1}}\right)^2 + \left(\Gamma_t \, \frac{m_t(m_{\tilde{t}_1}-m_t)}{m_{\tilde{t}_1}}\right)^2 } \, {\rm d}P \ .
\end{align}
For example, for $\Delta m=2$~GeV, $m_{\tilde{t}_1}=223$~GeV, $m_t=173$~GeV, $\Gamma_t=1.56$~GeV,
 $\bar{f}(P)$ is maximized at $P\simeq 13$~GeV;
 for $\Delta m=0$~GeV, $m_{\tilde{t}_1}=223$~GeV, $m_t=173$~GeV, $\Gamma_t=1.56$~GeV,
 $\bar{f}(P)$ is maximized at $P\simeq 8$~GeV.

In the frame where the mother \textit{stop} is boosted transversely by the 2-velocity $(\beta, 0)$,
 the transverse momenta of the reconstructed top quark and the neutralino are given by
\begin{align}
\vec{p}_{t \, T} \ &= \ ( \, \gamma \beta \sqrt{P^2+m_t^2}+\gamma P \cos \theta, \, P \sin \theta \cos \phi \, ), \nonumber \\
\vec{p}_{\tilde{\chi}^0_1 \, T} \ &= \ ( \, \gamma \beta \sqrt{P^2+m_{\tilde{\chi}^0_1}^2}-\gamma P \cos \theta, \, -P \sin \theta \cos \phi \, ),
\end{align}
 where $\gamma=1/\sqrt{1-\beta^2}$ and 
 ($P \cos \theta, \, P \sin \theta \cos \phi$) is the transverse component of the top quark momentum vector in the \textit{stop} rest frame.
We find that as the mother \textit{stop} is highly boosted with $\beta \rightarrow 1$,
 the momentum vectors $\vec{p}_{t \, T}$ and $\vec{p}_{\tilde{\chi}^0_1 \, T}$ become collinear
 while the ratio of their absolute values, $p_{\tilde{\chi}^0_1 \, T} / p_{t \, T}$,
 only approaches to
\begin{align*}
\frac{\sqrt{P^2+m_{\tilde{\chi}^0_1}^2}+ P \cos \theta}{\sqrt{P^2+m_t^2}+ P \cos \theta}
\end{align*}
 and never converges to $m_{\tilde{\chi}^0_1}/m_t$.
As for \textit{stop} pair production events,
 the departure of $p_{\tilde{\chi}^0_1 \, T} / p_{t \, T}$ from $m_{\tilde{\chi}^0_1}/m_t$
 gives rise to the deviation of $\Delta \phi( \vec{\mpt}_T, \, \vec{p}_{t\bar{t} \, T} )$ from zero.
However, we can alleviate such a deviation by selecting events with small azimuthal angle between the two top quarks.
On the other hand, the deviation of $\mpt_T/p_{t\bar{t} \, T}$ from $m_{\tilde{\chi}^0_1}/m_t$
 cannot be lessened by any event selection criteria.
To summarize, we have \textit{stop} events gathered near the line of 
 $\Delta \phi( \vec{\mpt}_T, \, \vec{p}_{t\bar{t} \, T} )=0$
 even with the sizable top quark decay width and even when the mass spectrum is slightly different from the one in the equal-velocity limit,
 if we select those events where the two reconstructed top quarks have large transverse momenta
 and small azimuthal angle separation.
Still, the \textit{stop} events are scattered along the axis of $\mpt_T/p_{t\bar{t} \, T}$
 and only possess a broad peak around the point of $\mpt_T/p_{t\bar{t} \, T}=m_{\tilde{\chi}^0_1}/m_t$.
\\

First we simulate \textit{stop} pair production events assuming an ideal detector with which top quark momenta are reconstructed perfectly
 and missing transverse momentum solely comes from neutralinos when the top quarks decay hadronically,
 and study the distribution of the simulated events on the plane of 
\begin{align*}
(\Delta \phi( \vec{\mpt}_T, \, \vec{p}_{t\bar{t} \, T} ), \, \mpt_T/p_{t\bar{t} \, T}) \ .
\end{align*}
We thereby examine to what extent \textit{stop} pair production events are concentrated around the point of
\begin{align*}
(\Delta \phi( \vec{\mpt}_T, \, \vec{p}_{t\bar{t} \, T} ), \, \mpt_T/p_{t\bar{t} \, T}) \, &= \, (0, \, m_{\tilde{\chi}^0_1}/m_t) \ .
\end{align*}
By using MadGraph5 and MadEvent \cite{mg} with the parton distribution function set NN23LO1 \cite{nn23lo1}, we calculate the matrix elements for the process eq.~(\ref{parton}) with $\sqrt{s}=14$~TeV,
\begin{align}
p p \ &\rightarrow \ \tilde{t}_1 \bar{\tilde{t}}_1 \ + \ 1 \ {\rm jet}, \ \ \
\tilde{t}_1 \ \rightarrow \ b \, j \, j \, \tilde{\chi}^0_1, \ \ \ 
\bar{\tilde{t}}_1 \ \rightarrow \ \bar{b} \, j \, j \, \tilde{\chi}^0_1
\label{parton}
\end{align}
 where $b$ denotes a bottom quark jet and $j$ denotes a light flavor jet,
 and thus generate parton-level events.
We here focus on events with the hadronic decay of the top quark pair,
 to realize full reconstruction of the top quark momenta and
 avoid contamination of missing transverse momentum from neutrinos due to top quark decays.
We require one additional jet along with the \textit{stop} pair in the event generation
 because in the equal-velocity limit, missing transverse momentum from the neutralinos 
 arises only when the \textit{stop} pair is boosted.
We test the following four benchmark mass spectra:
\begin{align*}
(m_{\tilde{t}_1}, \ m_{\tilde{\chi}^0_1}, \ \Delta m = m_{\tilde{t}_1}-m_t-m_{\tilde{\chi}^0_1})
 & \ = \ (223 {\rm GeV}, \ 50 {\rm GeV}, \ 0 {\rm GeV}) & {\rm (11a)} \\
 & \ = \ (223 {\rm GeV}, \ 48 {\rm GeV}, \ 2 {\rm GeV}) & {\rm (11b)} \\
 & \ = \ (323 {\rm GeV}, \ 150 {\rm GeV}, \ 0 {\rm GeV}) & {\rm (11c)} \\
 & \ = \ (323 {\rm GeV}, \ 148 {\rm GeV}, \ 2 {\rm GeV}) & {\rm (11d)}
\end{align*}
 with the top quark mass and width set at $m_t=173$~GeV and $\Gamma_t=1.56$~GeV.
For simplicity, it is assumed that the lighter \textit{stop} $\tilde{t}_1$ is purely composed of SU(2)$_L$ singlet \textit{stop}
 and the lightest neutralino $\tilde{\chi}^0_1$ is composed of Bino,
 and that the other SUSY particles are decoupled from the mass spectrum.
Hence the \textit{stop} decays into the neutralino and the top quark with 100\% branching ratio.
The \textit{stop} decay width is calculated and  
 found to be $\Gamma_{\tilde{t}_1}=$0.00753~GeV, 0.0183~GeV, 0.0145~GeV and 0.0371~GeV
 for the mass spectra eqs.~(11a), (11b), (11c) and (11d), respectively.
For events generated, we require that
\begin{itemize}
 \item the transverse momentum of each of the reconstructed top quarks should satisfy $p_T > 200$~GeV.  \ \ \ {\rm (12a)}
 \item the azimuthal angle between the reconstructed top quarks should satisfy $\Delta \phi < \pi/2$.   \ \ \ {\rm (12b)}
\end{itemize}
They reflect our observations mentioned in the previous paragraph.
\setcounter{equation}{12}

In Figure~1, we plot events of the process eq.~(\ref{parton})
 on the 2-dimensional plane spanned by the azimuthal angle between 
 the missing transverse momentum vector $\vec{\mpt}_T$ and the top-quark-pair transverse momentum vector $\vec{p}_{t\bar{t} \, T}$,
 and the ratio of their absolute values.
Here the top-quark-pair transverse momentum vector is reconstructed from the six jets coming from the hadronic decay of the top quark pair.
The four subfigures, (1a) to (1d), show results for the four mass spectra, eqs.~(11a) to (11d), respectively.
The leading order perturbative QCD calculation predicts the cross section for the process eq.~(\ref{parton})
 after the selection cuts eqs.~(12a) and (12b) to be 
 30.9~fb for $(m_{\tilde{t}_1}, \ m_{\tilde{\chi}^0_1})=$(223~GeV, 50~GeV),
 33.6~fb for $(m_{\tilde{t}_1}, \ m_{\tilde{\chi}^0_1})=$(223~GeV, 48~GeV),
 2.71~fb for $(m_{\tilde{t}_1}, \ m_{\tilde{\chi}^0_1})=$(323~GeV, 150~GeV),
 3.19~fb for $(m_{\tilde{t}_1}, \ m_{\tilde{\chi}^0_1})=$(323~GeV, 148~GeV),
 at $\sqrt{s}=14$~TeV.
Note that the cross sections do depend on $m_{\tilde{\chi}^0_1}$ because the selection cut eq.~(12b) is sensitive to
 the proportion of the on-shell top quark contribution to the \textit{stop} decay.
Each dot corresponds to 1~event for 10~fb$^{-1}$ of data in subfigures~(1a) and (1b),
 and to 1~event for 100~fb$^{-1}$ of data in subfigures~(1c) and (1d).
\begin{figure}[htbp]
 \begin{minipage}{0.5\hsize}
  \begin{center}
   \includegraphics[width=80mm]{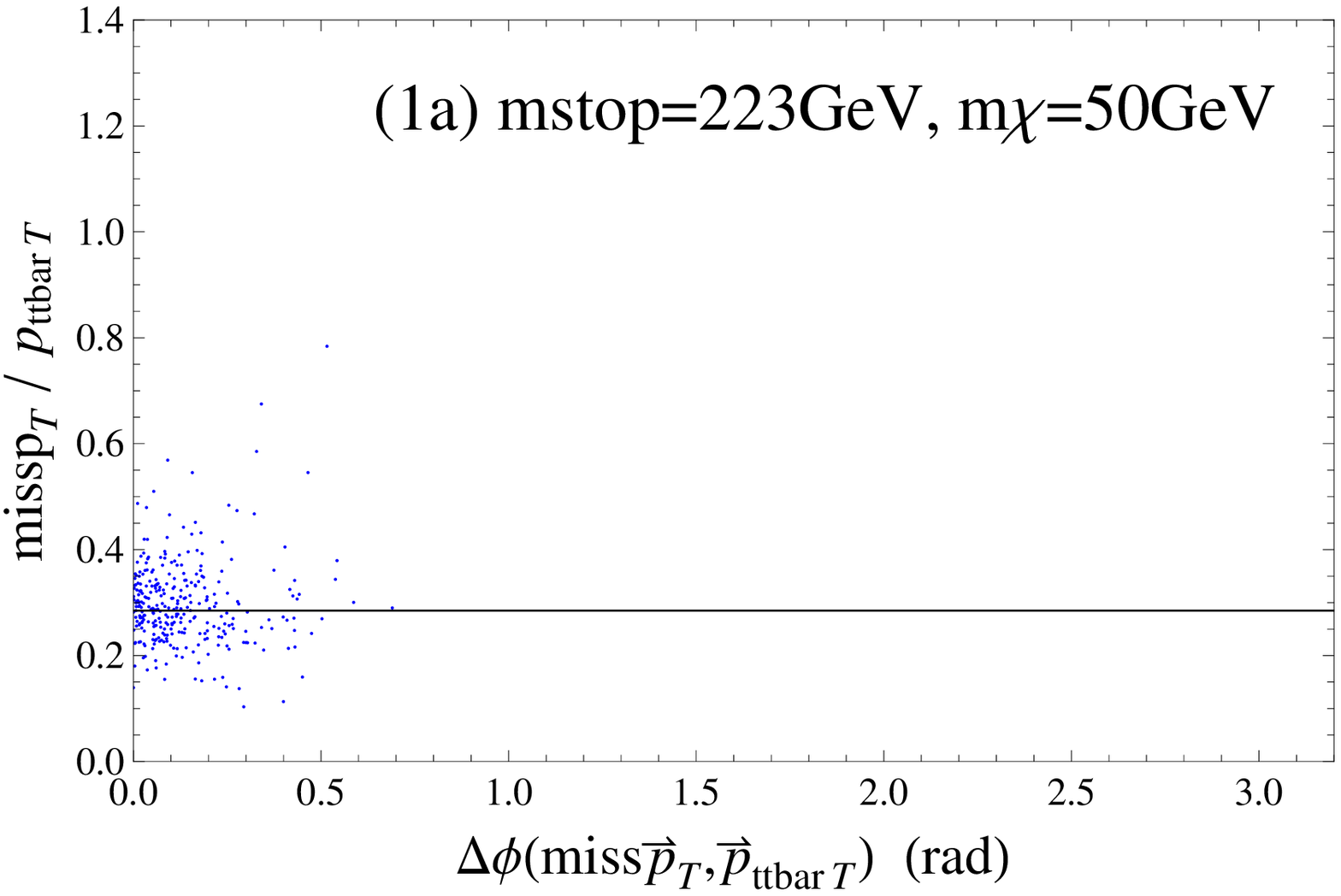}
  \end{center}
 \end{minipage}
 \begin{minipage}{0.5\hsize}
  \begin{center}
   \includegraphics[width=80mm]{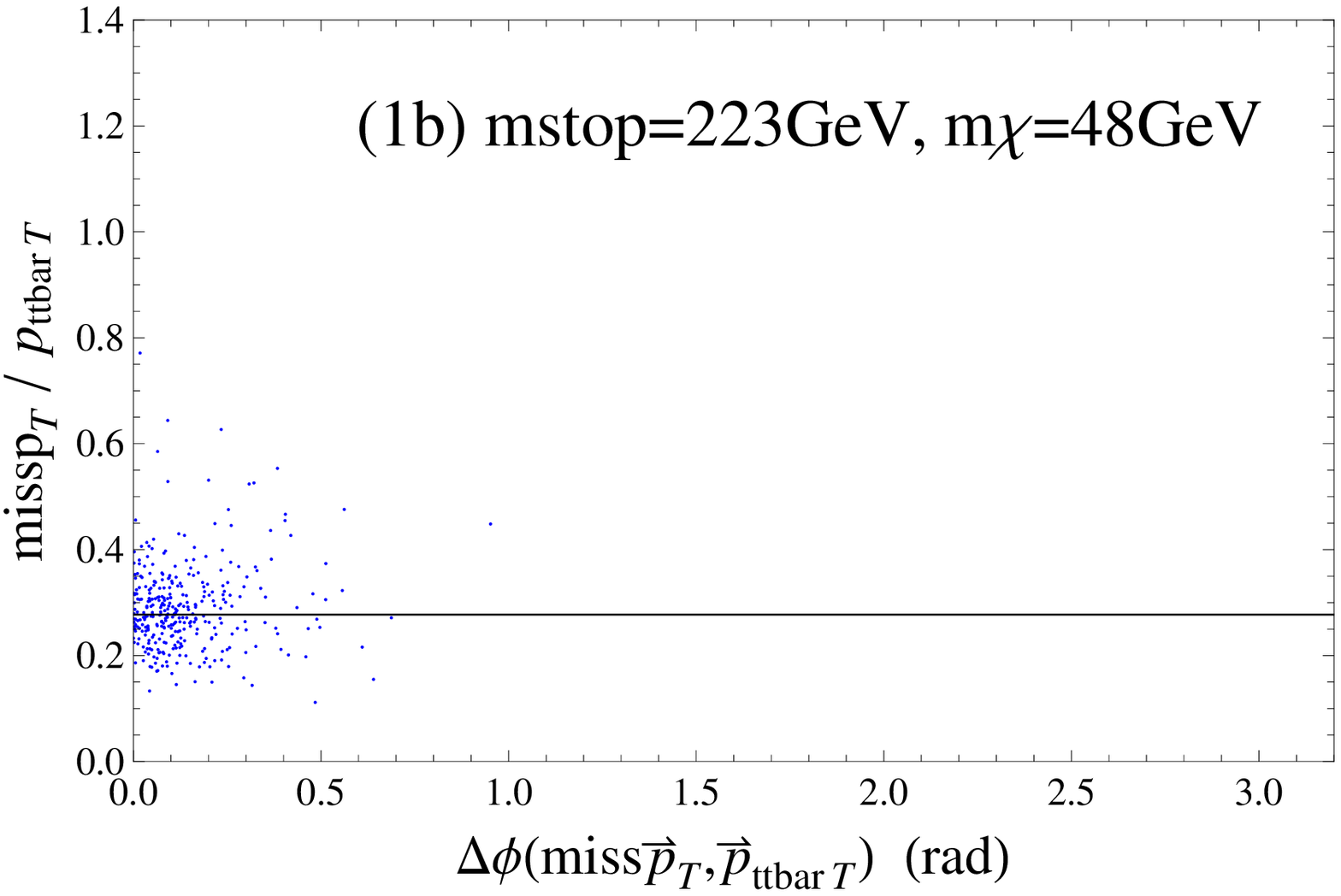}
  \end{center}
 \end{minipage}
 \begin{minipage}{0.5\hsize}
  \begin{center}
   \includegraphics[width=80mm]{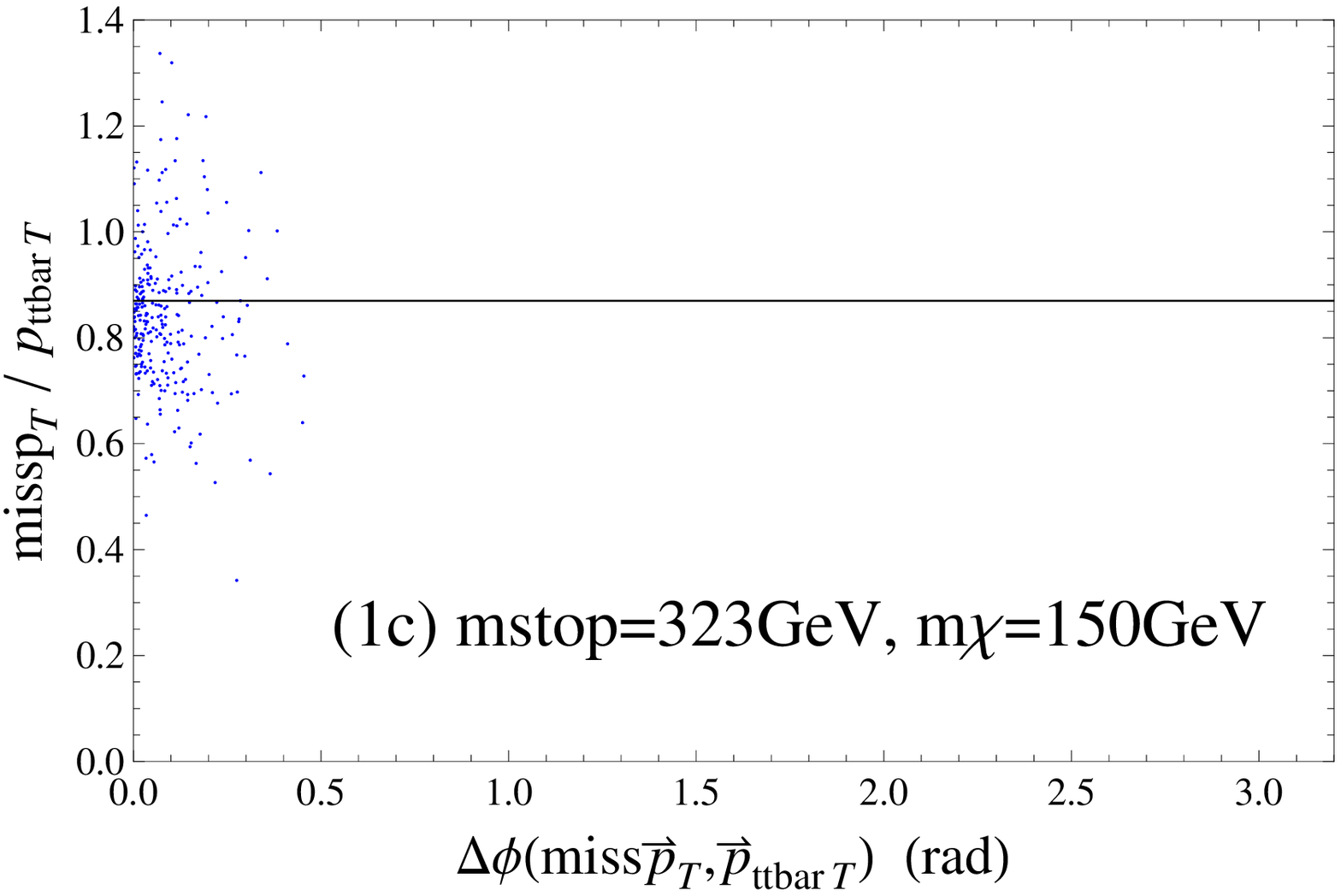}
  \end{center}
 \end{minipage}
 \begin{minipage}{0.5\hsize}
  \begin{center}
   \includegraphics[width=80mm]{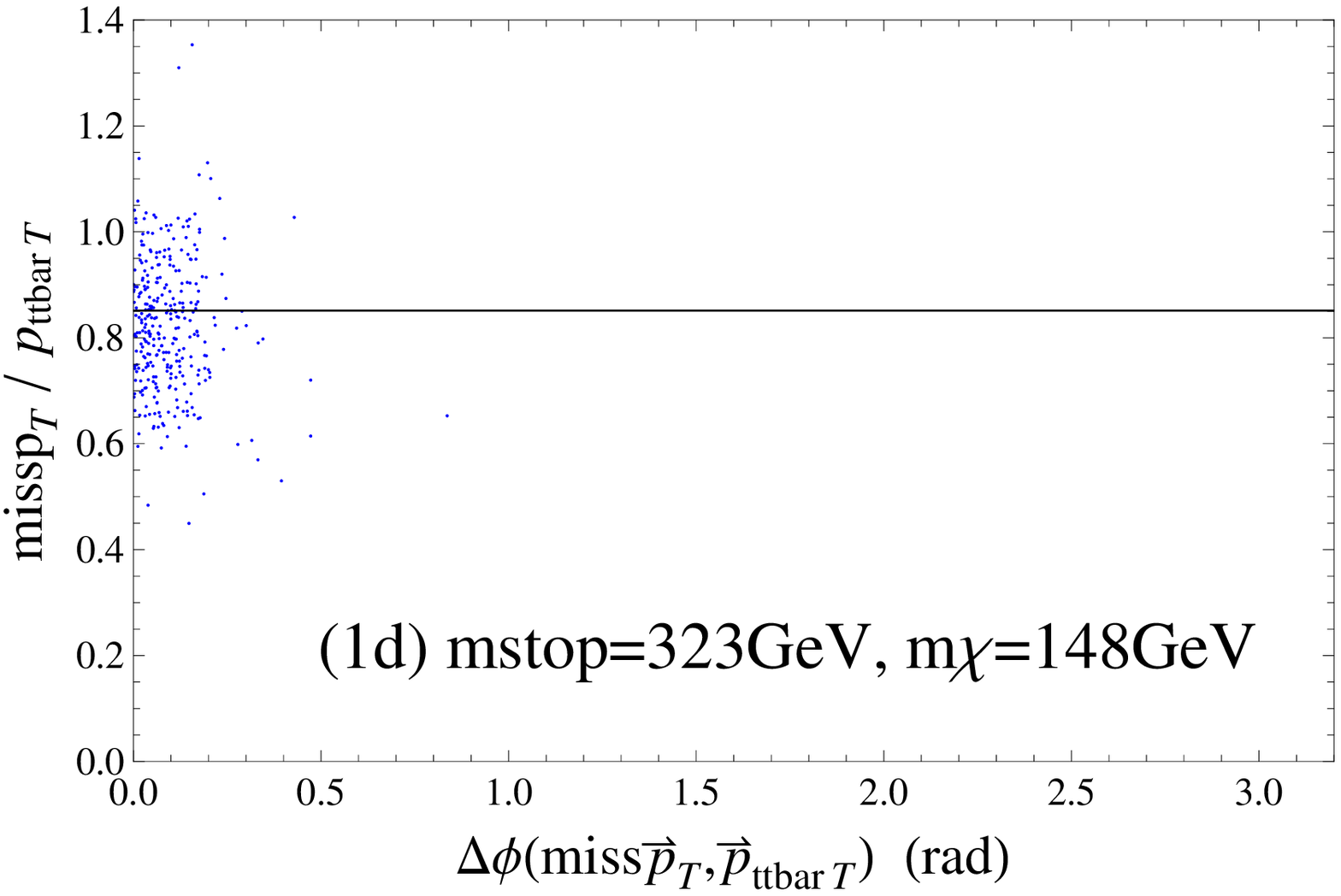}
  \end{center}
 \end{minipage}
 \caption{The scatter plots of events of the process eq.~(\ref{parton}) after the selection cuts eqs.~(12a) and (12b) 
  in $\sqrt{s}=14$~TeV $pp$ collisions,
  on the plane spanned by the azimuthal angle between the missing transverse momentum vector $\vec{\mpt}_T$
  and the top-pair transverse momentum vector $\vec{p}_{t\bar{t} \, T}$,
  $\Delta \phi (\vec{\mpt}_T, \vec{p}_{t\bar{t} \, T})$,
  and the ratio of their absolute values,
  $\mpt_T / p_{t\bar{t} \, T}$.
 It is assumed that the top-quark-pair momentum is perfectly reconstructed and that missing transverse momentum solely
  comes from neutralinos.
 The four subfigures are for
  $(m_{\tilde{t}_1}, \, m_{\tilde{\chi}^0_1}) =$(223~GeV,~50~GeV) (1a),
  (223~GeV,~48~GeV) (1b), (323~GeV,~150~GeV) (1c) and (323~GeV,~148~GeV) (1d),
  respectively.
 The thick horizontal line inside each figure corresponds to the line of $\mpt_T / p_{t\bar{t} \, T} = m_{\tilde{\chi}^0_1}/m_t$.
 }
\end{figure}
From Figure~1, we have confirmed that the \textit{stop} pair production events gather around the point of
 ($\Delta \phi( \vec{\mpt}_T, \, \vec{p}_{t\bar{t} \, T} ), \, \mpt_T/p_{t\bar{t} \, T}) \, = \, (0, \, m_{\tilde{\chi}^0_1}/m_t)$.
Most of the \textit{stop} events are distributed in the region with $\Delta \phi( \vec{\mpt}_T, \, \vec{p}_{t\bar{t} \, T} )<\pi/10$,
 which suggests that $\Delta \phi( \vec{\mpt}_T, \, \vec{p}_{t\bar{t} \, T} )$ can be an efficient discriminating variable
 for selecting \textit{stop} events in the presence of SM backgrounds.
On the other hand, the event distribution along the axis of $\mpt_T/p_{t\bar{t} \, T}$ can be useful for estimating the mass of the \textit{stop} and neutralino.

In subfigures~(1a) and (1c), the events are scattered due to the finite width of the top quark,
 whereas in subfigures~(1b) and (1d), they are scattered due to the combination of the finite top quark width
 and the deviation of the mass spectrum from the equal-velocity limit by $\Delta m=2$~GeV.
As a reference, we have also checked the event distributions for the mass spectra eqs.~(1b) and (1c) with the top quark width set to zero,
 namely, with the assumption that only on-shell top quark contributes to the \textit{stop} decay.
Such distributions are unphysical, but useful for studying how a finite value of $\Delta m \equiv m_{\tilde{t}_1} - m_t - m_{\tilde{\chi}^0_1}$
 alone causes the scattering of events.
It turns out that the distributions of \textit{stop} events on 
 the plane of ($\Delta \phi( \vec{\mpt}_T, \, \vec{p}_{t\bar{t} \, T} ), \, \mpt_T/p_{t\bar{t} \, T}$)
 are similar for the cases with $(\Delta m, \, \Gamma_t)=(0, \, 1.56)$~GeV, $(\Delta m, \, \Gamma_t)=(2, \, 1.56)$~GeV and 
 $(\Delta m, \, \Gamma_t)=(2, \, 0)$~GeV.
Based on this observation, in the detector-level analysis later on, 
 we neglect off-shell top quark contribution and consider mass spectra with $(\Delta m, \, \Gamma_t)=(2, \, 0)$~GeV.
\\

Next we perform a detector-level analysis, taking into account parton showering, hadronization and detector responses.
We again focus on \textit{stop} events with the hadronic decay of the top quark pair, where
 the momenta of the top quarks can be reconstructed from six jets.
In addition, we consider the dominant background that originates from SM top quark pair production events.
We examine how effective the plot on the plane of 
 ($\Delta \phi( \vec{\mpt}_T, \, \vec{p}_{t\bar{t} \, T} ), \, \mpt_T/p_{t\bar{t} \, T}$) is
 for distinguishing \textit{stop} events from SM backgrounds in the equal-velocity scenario.
The benchmark mass spectra are eqs.~(11b) and (11d), and we take into account only on-shell top quark contribution to
 the \textit{stop} decay, which has been justified in the previous paragraph.
It is again assumed
 that the lighter \textit{stop} $\tilde{t}_1$ is purely composed of SU(2)$_L$ singlet \textit{stop},
 the lightest neutralino $\tilde{\chi}^0_1$ is composed of Bino and
 the other SUSY particles are decoupled.

Before event simulation, we evaluate the cross sections for the \textit{stop} pair production and the SM top quark pair production
 to assess the viability of the analysis.
In Figure~2, the cross sections for the \textit{stop} pair production, $\sigma(\tilde{t}_1 \bar{\tilde{t}}_1)$,
 the \textit{stop} pair production associated with
 one parton whose transverse momentum $p_{jT}$ is larger than 300~GeV,
 $\sigma(\tilde{t}_1 \bar{\tilde{t}}_1 j; p_{jT}>300 {\rm GeV})$,
 and the \textit{stop} pair production associated with
 one parton with $p_{jT}>$500~GeV,
 $\sigma(\tilde{t}_1 \bar{\tilde{t}}_1 j; p_{jT}>500 {\rm GeV})$,
 in $\sqrt{s}=14$~TeV $pp$ collisions are plotted.
These cross sections are calculated at the leading order in perturbative QCD.
Also shown in Figure~2 are the cross sections for the SM top quark pair prodcution, $\sigma(t \bar{t})_{LO}$, $\sigma(t \bar{t})_{NLO}$,
 and the SM top quark pair production associated with
 one parton with $p_{jT}>300$~GeV, $\sigma(t \bar{t}; p_{jT}>300 {\rm GeV})_{LO}$, $\sigma(t \bar{t}; p_{jT}>300 {\rm GeV})_{NLO}$,
 which are calculated at the leading order for $\sigma(t \bar{t})_{LO}$, $\sigma(t \bar{t}; p_{jT}>300 {\rm GeV})_{LO}$
 and at the next-to-leading order for $\sigma(t \bar{t})_{NLO}$, $\sigma(t \bar{t}; p_{jT}>300 {\rm GeV})_{NLO}$ in perturbative QCD.
Here the next-to-leading order calculation is done with MadGraph5$\underline{ \ }$aMC@NLO \cite{mg}.
The K-factor for the SM top quark pair production cross section is evaluated to be 1.34,
 and that for the production cross section of a SM top quark pair + one parton with $p_{jT}>300$~GeV 
 is evaluated to be 1.09.
The process of \textit{stop} pair production associated with one parton with $p_{jT}>300$~GeV
 resembles the situation where a pair of \textit{stop}s with the mass of 223~GeV are produced with a transverse boost balanced by a hard jet,
 and each of the top quarks coming from their decays has a transverse momentum above 200~GeV
 and their azimuthal angle is below $\pi/2$, which is the situation where the selection cuts eqs.~(16g) and (16h) below are imposed.
Likewise, the process of \textit{stop} pair production associated with one parton with $p_{jT}>500$~GeV
 resembles the situation where a pair of \textit{stop}s with the mass of 323~GeV are produced with the same selection cuts.
The process of top quark pair production associated with one parton with $p_{jT}>300$~GeV
 resembles the situation where a top quark pair is produced with a transverse boost balanced by a hard jet,
 with the transverse momentum of each top quark above 200~GeV
 and with their azimuthal angle below $\pi/2$.
\begin{figure}[htbp]
  \begin{center}
   \includegraphics[width=100mm]{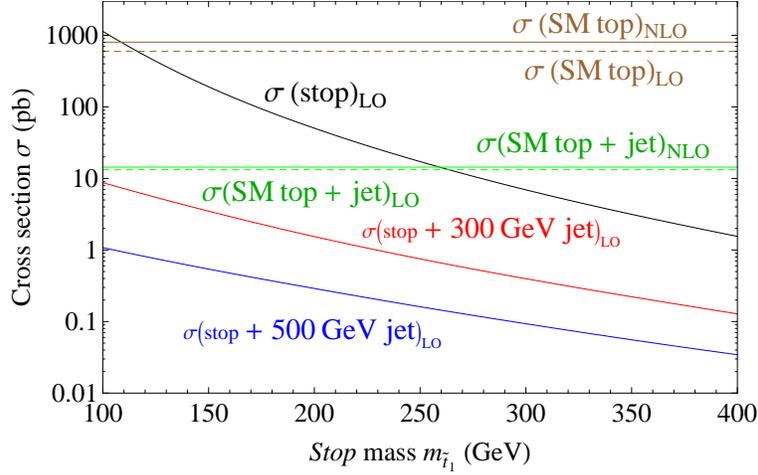}
  \end{center}
 \caption{The production cross section of a pair of \textit{stop}s of mass $m_{\tilde{t}_1}$
  (black upper solid line), that of a pair of \textit{stop}s associated with one parton
  whose transverse momentum is above 300~GeV (red middle solid line), 
  and that of a pair of \textit{stop}s associated with one parton
  whose transverse momentum is above 500~GeV (blue lower solid line), 
  in $\sqrt{s}=14$~TeV $pp$ collisions,
  calculated at the leading order in perturbative QCD.
 Also shown are the production cross section of a SM top quark pair,
  and that of a SM top quark pair associated with one parton
  whose transverse momentum is above 300~GeV.
 For the top quark pair production, the cross sections are calculated both at the leading order
  and at the next-to-leading order in perturbative QCD.
 The brown upper and green lower dashed lines correspond to
  the top quark pair production cross sections calculated at the leading order and the next-to-leading order,
  respectively.
 The brown upper and green lower solid lines correspond to
  the production cross sections for a top quark pair plus one jet with $p_T>300$~GeV,
  calculated at the leading order and the next-to-leading order,
  respectively.
 }
\end{figure}

Let us proceed to event simulation with full detector-level analysis.
By MadGraph5 and MadEvent with the parton distribution function set NN23LO1,
 we generate signal events from the following processes
\begin{align}
p p \ &\rightarrow \ \tilde{t}_1 \bar{\tilde{t}}_1 \ + \ 1 \ {\rm jet}, \ \ 
\tilde{t}_1 \ \rightarrow \ t \, \tilde{\chi}^0_1, \ \ 
\bar{\tilde{t}}_1 \ \rightarrow \ \bar{t} \, \tilde{\chi}^0_1; \nonumber
\\
p p \ &\rightarrow \ \tilde{t}_1 \bar{\tilde{t}}_1 \ + \ 2 \ {\rm jets}, \ \ 
\tilde{t}_1 \ \rightarrow \ t \, \tilde{\chi}^0_1, \ \ 
\bar{\tilde{t}}_1 \ \rightarrow \ \bar{t} \, \tilde{\chi}^0_1,
\label{sig}
\end{align}
 with the precuts that
\begin{itemize}
 \item each top quark should have a transverse momentum above 150~GeV. \hspace{\fill} (14a)
 \item the azimuthal angle between the two top quarks shoule be smaller than $2\pi/3$. \hspace{\fill} (14b)
 \item the $k_T$ distance of the jets produced in association with the \textit{stop} pair
  should be above 100~GeV. \hspace{\fill} (14c)
\end{itemize}
\setcounter{equation}{14}
Parton showering and hadronization are simulated by PYTHIA8 \cite{pythia},
 and the 1-jet and 2-jet events are matched by MLM matching scheme \cite{mlm}
 with the matching scale of 100~GeV.
The top quark decay is simulated by PYTHIA8,
 with the off-shell top quark contribution to the \textit{stop} decay neglected.
Detector simulation and jet clustering are performed by PGS4 \cite{pgs},
 where we use anti-$k_T$ algorithm \cite{antikt} with the distance parameter of $\Delta R=0.4$.
We also generate background events from the following processes by MadGraph5 and MadEvent with the same precuts,
\begin{align}
p p \ &\rightarrow \ t \bar{t} \ + \ 1 \ {\rm jet}; \ \ \nonumber
\\
p p \ &\rightarrow \ t \bar{t} \ + \ 2 \ {\rm jets}. \ \ 
\label{bkg}
\end{align}
Parton showering, hadronization, the top qurak decay, event matching, detector responses and jet clustering 
 are simulated in the same way as the signal events.
The next-to-leading order QCD corrections are taken into account only for the background process,
 by multiplying the number of background events by the K-factor evaluated in Figure~2
 for the production of a SM top pair + one parton with $p_T>300$~GeV, which has been found to be 1.09.

We impose the following selection cuts on the signal and background events generated:
\begin{itemize}
 \item Require six jets whose pseudo-rapidities satisfy $\vert \eta_j \vert < 2.8$
  and whose transverse momenta satisfy $p_{jT} > 30$~GeV. \hspace{\fill} (16a)
 \item Exactly two jets should be b-tagged. \hspace{\fill} (16b)
 \item Require missing transverse momentum $\mpt_T > 30$~GeV. \hspace{\fill} (16c)
 \item Azimuthal angle between the missing transverse momentum vector and the transverse momentum vector
  of each of the three leading jets should be larger than $0.1\pi$,
  
  $\Delta \phi(\vec{\mpt}_T, \, \vec{p}_{j_{1,2,3}T}) > 0.1\pi$. \hspace{\fill} (16d)
 \item The event is vetoed if it contains an electron with $\vert \eta \vert < 2.47$ and $p_T > 10$~GeV,
 or a muon with $\vert \eta \vert < 2.4$ and $p_T > 10$~GeV. \hspace{\fill} (16e)
 \item Hadronic decays of two top quarks should be reconstructed successfully by the method below. \hspace{\fill} (16f)
 \item The reconstructed top quarks, $t_1, t_2$, should satisfy 
 
 $p_{t_1 \, T} > 200$~GeV, $p_{t_2 \, T} > 200$~GeV. \hspace{\fill} (16g)
 \item The azimuthal angle between the reconstructed top quarks should satisfy 
   
 $\Delta \phi(\vec{p}_{t1 \, T}, \, \vec{p}_{t2 \, T}) < 0.5\pi$. \hspace{\fill} (16h)
\end{itemize}
In eq.~(16f), we use the following method to reconstruct the two top quarks that decay hadronically.
First we select a pair of non-b-tagged jets, $j_1, \, j_2,$ whose invariant mass is most close to the $W$ boson mass.
If the invariant mass is separated from the $W$ boson mass by more than 20~GeV, the method fails and the event is vetoed:
\begin{align*}
{\rm Look \ for \ } j_1, \, j_2 \ {\rm that \ give \ the \ minimum \ of} \ \vert m(j_1, \, j_2) - M_W \vert.
\\ {\rm The \ event \ is \ vetoed \ if} \ {\rm min}\{ \vert m(j_1, \, j_2) - M_W \vert \} \, > \, 20 \, {\rm GeV}.
\end{align*}
Next we select one b-tagged jet, $b_1$, for which the three-jet invariant mass for this jet and the two non-b-tagged jets selected above, $\bar{j}_1, \, \bar{j}_2,$
 is most close to the top quark mass.
If the invariant mass is separated from the top quark mass by more than 50~GeV, the method fails and the event is vetoed:
\begin{align*}
{\rm Look \ for \ } b_1 \ {\rm that \ gives \ the \ minimum \ of} \ \vert m(\bar{j}_1, \, \bar{j}_2, \, b_1) - m_t \vert.
\\ {\rm The \ event \ is \ vetoed \ if} \ {\rm min}\{ \vert m(\bar{j}_1, \, \bar{j}_2, \, b_1) - m_t \vert \} \, > \, 50 \, {\rm GeV}.
\end{align*}
The two non-b-tagged jets $j_1, \, j_2$ and one b-tagged jet $b_1$ are considered to constitute one hadronically decaying top quark.
We repeat the same procedures on the remaining jets to reconstruct the other top quark.
This method can be regarded as a simplified version of the one used in ref.~\cite{atlas sm hadronic top},
 where a likelihood function is utilized to select two sets of two non-b-tagged jets and one b-tagged jet that constitute a hadronically decaying top quark.
Note that in our method, we do not rely on information about the angular distance between a pair of jets or combined jets
This is because the decay products of the two top quarks are likely to overlap in the $\phi-\eta$ space,
 as eq.~(16h) demands that the two top quarks have small azimuthal angular distance.
\setcounter{equation}{16}

Our simulation of b-tagging in eq.~(16b) is based on "loose-b-tag" implemented in PGS4 \cite{pgs},
 which gives a conservative estimate on the power of b-tagging compared to the performance achieved by the ATLAS and CMS collaborations.
Here the efficiency and mis-tagging rate are simulated to depend on the energy and pseudo-rapidity of the jet.
The efficiency of b-tagging is simulated as above 40\% and below 60\% for b-jets with $p_T > 20$~GeV and $\vert \eta_j \vert < 1.0$.
The rate for mis-tagging a light flavor jet coming from up, down or strange quark or gluon is simulated to be suppressed below 4\%,
 and that for mis-tagging a charm jet is set to be 26\% of the b-tagging efficiency.

We have further comments on the selection cuts.
The mild requirement on $\mpt_T$ in eq.~(16c) is because
 missing transverse momentum is suppressed in the equal-velocity scenario compared to the case with general mass spectra.
The cut eq.~(16d) is to reduce fake missing transverse momentum due to jet energy mismeasurement.
The cut eq.~(16e) is to reduce SM top quark pair background events where missing transverse momentum
 originates from their semi-leptonic decays.
The cuts eqs.~(16g) and (16h) are special cuts that ameliorate the concentration of \textit{stop} signal events
 on the plane of ($\Delta \phi( \vec{\mpt}_T, \, \vec{p}_{t\bar{t} \, T} ), \, \mpt_T/p_{t\bar{t} \, T}$).

We find that the cross section of the \textit{stop} pair production process
 after the selection cuts eqs.~(16a)-(16h)
 is 0.60~fb and 0.077~fb
 for the mass spectra $(m_{\tilde{t}_1}, \, m_{\tilde{\chi}_1^0})=$(223,~48)~GeV and 
 (323,~148)~GeV, respectively,
 and the cross section of the SM top quark pair production process after the same cuts
 is 5.95~fb.
Here, the K-factor of 1.09 derived in Figure~2 has already been multiplied to the top quark pair production cross section.
Table~1 shows the cross sections of \textit{stop} pair production events with 
 $(m_{\tilde{t}_1}, \, m_{\tilde{\chi}_1^0})=$(223,~48)~GeV and (323,~148)~GeV
 and top quark pair production events, at each stage of event selection.
\begin{table}
\begin{center}
\begin{tabular}{|c|c|c|c|} \hline
 & $(m_{\tilde{t}_1}, \, m_{\tilde{\chi}_1^0})=$(223,~48)~GeV & $(m_{\tilde{t}_1}, \, m_{\tilde{\chi}_1^0})=$(323,~148)~GeV & SM top quark \\ \hline
Before the cuts           & 3.8 $\times 10^4$~fb & 6.6 $\times 10^3$~fb & 7.5 $\times 10^5$~fb \\ \hline
After eqs.~(16a,b,e,f)    & 4.2 $\times 10^2$~fb & 1.0 $\times 10^2$~fb & 9.1 $\times 10^3$~fb \\ \hline
After eqs.~(16c,d)        & 1.9 $\times 10^2$~fb & 63~fb                & 1.3 $\times 10^3$~fb \\ \hline
After eqs.~(16g,h)        & 0.60~fb              & 0.077~fb             & 5.95~fb \\ \hline
\end{tabular}
\end{center}
\caption{The cross sections of \textit{stop} pair production events with 
 $(m_{\tilde{t}_1}, \, m_{\tilde{\chi}_1^0})=$(223,~48)~GeV and (323,~148)~GeV
 and top quark pair production events at four stages of event selection that are,
 before the cuts, after the reconstruction of a hadronically-decaying top quark pair eqs.~(16a,~16b,~16e,~16f),
 after the cuts on missing transverse momentum eqs.~(16c,~16d), and after the cuts on the momenta of reconstructed top quarks eqs.~(16g,~16h).
}
\end{table}
Note that in Table~1, the cross sections before the selection cuts eqs.~(16g, 16h) are evaluated
 by generating \textit{stop} pair or top quark pair production events with 0~jet and 1~jet with
 the precut of eq.~(14c) only, passing them to parton showering and matching, and doing the rest of the simulations.
Also, the K-factor of 1.34, instead of 1.09, derived in Figure~2 has been multiplied to the top quark pair production cross sections
 before the cuts eqs.~(16g, 16h).

We plot in Figure~3 \textit{stop} signal events and SM top quark background events
 on the plane spanned by the azimuthal angle between missing transverse momentum vector $\vec{\mpt}_T$
 and the reconstructed top quark pair transverse momentum vector $\vec{p}_{t\bar{t} \, T}$,
 $\Delta \phi(\vec{p}_{t\bar{t} \, T}, \vec{\mpt}_T)$,
 and the ratio of their absolute values,
 $\mpt_T / p_{t\bar{t} \, T}$.
\begin{figure}[htbp]
  \begin{center}
   \includegraphics[width=90mm]{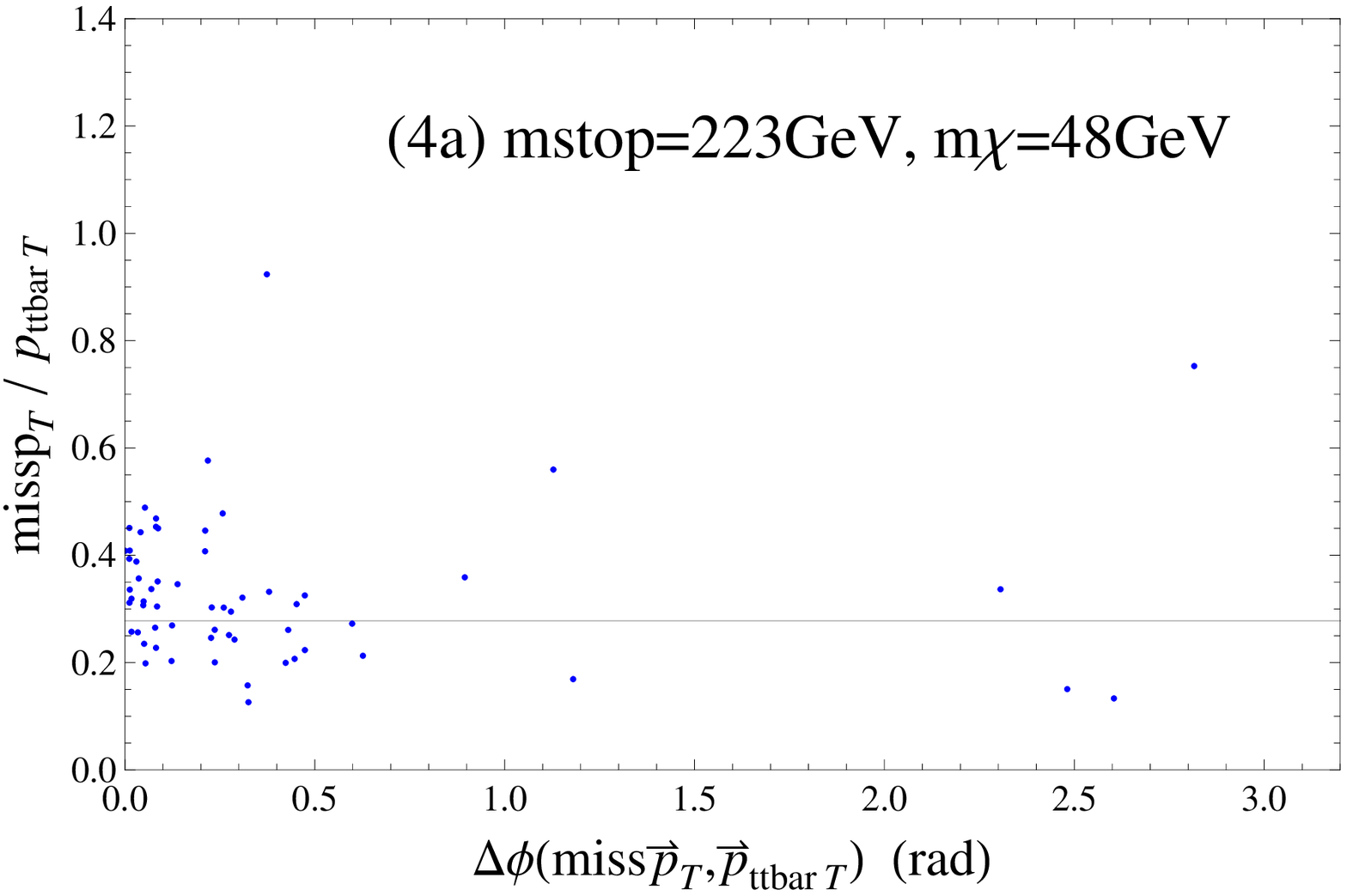}
  \end{center}
  \begin{center}
   \includegraphics[width=90mm]{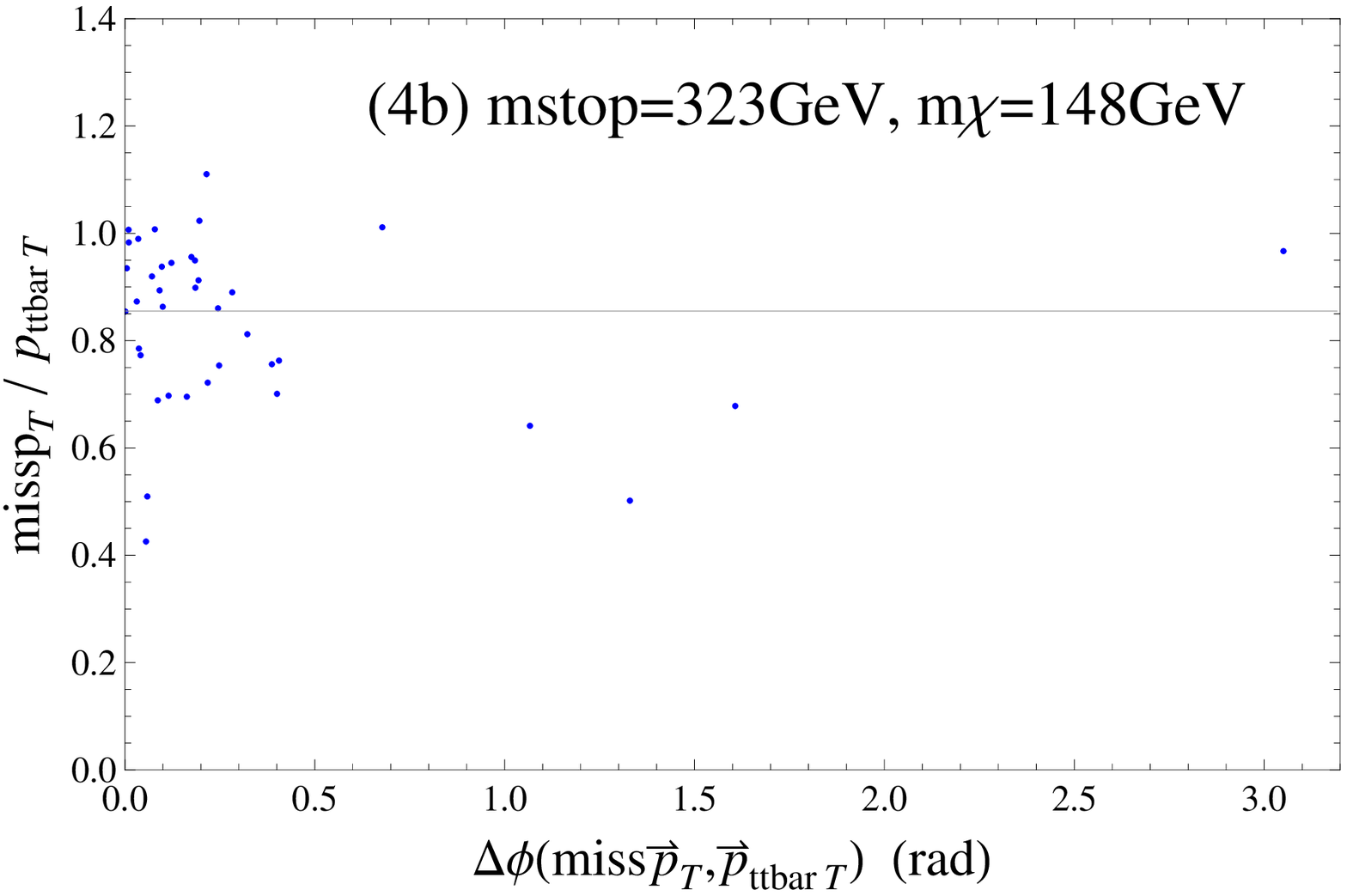}
  \end{center}
  \begin{center}
   \includegraphics[width=90mm]{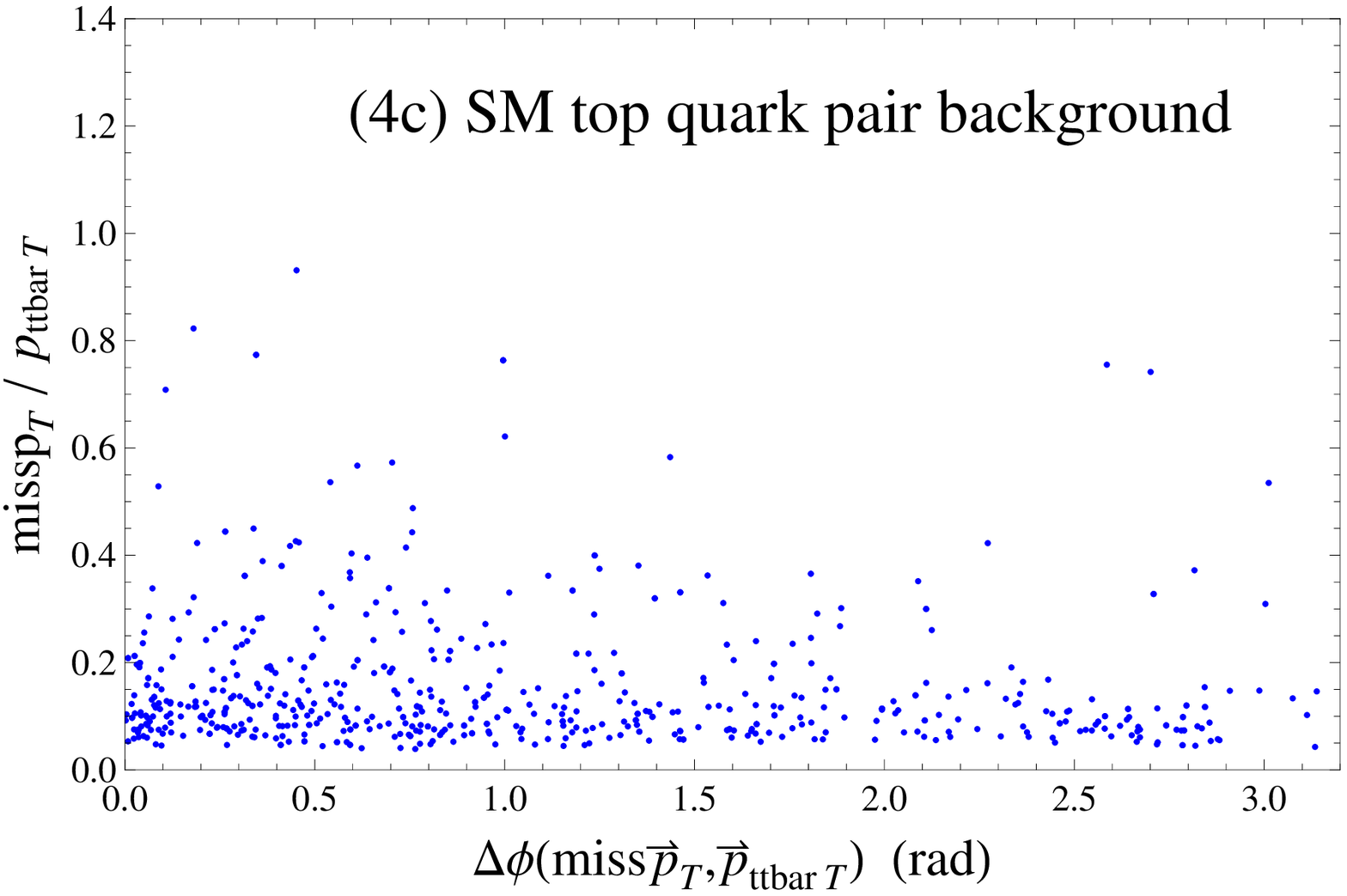}
  \end{center}
\caption{The scatter plots of \textit{stop} pair production events and SM top quark pair production events
  at the $\sqrt{s}=14$ TeV LHC that pass the selection cuts eqs.~(16a-16h),
  on the plane spanned by the azimuthal angle between the missing transverse momentum vector $\vec{\mpt}_T$
  and the top-pair transverse momentum vector $\vec{p}_{t\bar{t} \, T}$,
  $\Delta \phi (\vec{\mpt}_T, \vec{p}_{t\bar{t} \, T})$,
  and the ratio of their absolute values,
  $\mpt_T / \vec{p}_{t\bar{t} \, T}$.
 Subfigure~(4a) displays \textit{stop} pair production events with the mass spectrum of
  $(m_{\tilde{t}_1}, \, m_{\tilde{\chi}^0_1}) =$(223~GeV, 48~GeV),
  Subfigure~(4b) shows those with (323~GeV, 148~GeV),
  and Subfigure~(4c) shows SM top quark pair production events, which give the dominant contribution
  to the background.
 In Subfigures~(4a) and (4c), one dot corresponds to 1~event with 100~fb$^{-1}$ of data,
  while in Subfigure~(4b), one dot corresponds to 1~event with 500~fb$^{-1}$ of data.
 The thick horizontal lines inside Subfigures~(4a) and (4b) correspond to the lines of 
  $\mpt_T / \vec{p}_{t\bar{t} \, T}=m_{\tilde{\chi}^0_1}/m_t$.
 }
\end{figure}
From Figure~3, we find that \textit{stop} pair production events in the equal-velocity scenario 
 are concentrated around the point of
 $(\Delta \phi( \vec{\mpt}_T, \, \vec{p}_{t\bar{t} \, T} ), \, \mpt_T/p_{t\bar{t} \, T}) \, = \, 
 (0, \, m_{\tilde{\chi}^0_1}/m_t)$ even at the detector level.
We thus confirm that the kinematic relations of the top quark pair momentum and the missing transverse momentum
 which we observe at the parton-level in Figure~1
 are preserved through parton showering, hadronization, detector simulations, jet clustering
 and the reconstruction of hadronically-decaying top quarks.
On the other hand, SM top pair production background events are distributed broadly along the axis of 
 $\Delta \phi( \vec{\mpt}_T, \, \vec{p}_{t\bar{t} \, T} )$ and concentrated in the region with small
 $\mpt_T/p_{t\bar{t} \, T}$.
Such a distribution is explained in the following manner.
There are two principal sources of missing transverse momentum in the background events, which are
 the semi-leptonic decay of top quark pair where the charged lepton is a hadronically-decaying tau or evades detection,
 and the leptonic decay of B meson formed by a b-quark from the top quark decay.
The direction of such missing transverse momentum has no strong correlation with that of the momentum of wrongly-reconstructed 
 hadronically-decaying top quark pair, for which an initial or final state radiation is misidentified as a top quark decay product.
Also, the absolute value of such missing transverse momentum remains small 
 no matter how boosted the wrongly-reconstructed top quark pair is.
Comparison of Subfigures~(4a) and (4b) with Subfigure~(4c) implies that
 the selection cut on $\Delta \phi( \vec{\mpt}_T, \, \vec{p}_{t\bar{t} \, T} )$
 after the selection cuts eqs.~(16a-16h) is useful for enhancing the signal-to-background ratio
 in the search for \textit{stop}s in the equal-velocity scenario.
Then the distribution of $\mpt_T/p_{t\bar{t} \, T}$ gives a further criterion for distinguishing
 \textit{stop} events from SM top quark events if the neutralino mass is larger than about 50~GeV.
As a bonus, the same distribution makes it possible to estimate the neutralino mass and the \textit{stop} mass
 with the assumption of the equal-velocity scenario.

We finally demonstrate how the analysis using the scatter plots of Figure~3
 has an advantage over the analysis without the plots,
 by deriving the signal significance, $S/\sqrt{S+B}$, by both analyses.
In the latter analysis, we simply count the numbers of simulated signal and background events that pass the cuts eqs.~(16a-16h).
With 300~fb$^{-1}$ of data, the significance in the cases with 
 $(m_{\tilde{t}_1}, \, m_{\tilde{\chi}_1^0})=$(223,~48)~GeV and (323,~148)~GeV is found to be
\begin{align}
S/\sqrt{S+B} \ &= \ 4.1 \ \ \ {\rm for} \ (m_{\tilde{t}_1}, \, m_{\tilde{\chi}_1^0})=(223,~48)~{\rm GeV},
\nonumber \\
S/\sqrt{S+B} \ &= \ 0.54 \ \ \ {\rm for} \ (m_{\tilde{t}_1}, \, m_{\tilde{\chi}_1^0})=(323,~148)~{\rm GeV},
 \ {\rm with \ } 300~{\rm fb}^{-1} \ {\rm of \ data.} 
\end{align}
On the other hand, in the analysis that makes use of the scatter plots,
 we impose the selection cuts eq.~(18a, 18b) in addition to eqs.~(16a-16h):
\begin{align*}
\Delta \phi( \vec{\mpt}_T, \, \vec{p}_{t\bar{t} \, T} ) \ &< \ 0.5. \ \ \ \ \ \ \ \ \ \ (18a)
\\
\mpt_T/p_{t\bar{t} \, T} \ &> \ 0.2. \ \ \ \ \ \ \ \ \ \ (18b)
\end{align*}
\setcounter{equation}{18}
These cuts are based on the scatter plots of Figure~3, which indicate that
 the region defined by eqs.~(18a, 18b) is rich with signal events for cases with $m_{\tilde{\chi}_1^0}\gtrsim 50$~GeV,
 and contains considerably less background events than the region outside.
After the cuts eqs.~(18a, 18b), we find
\begin{align}
S/\sqrt{S+B} \ &= \ 8.4 \ \ \ {\rm for} \ (m_{\tilde{t}_1}, \, m_{\tilde{\chi}_1^0})=(223,~48)~{\rm GeV},
\nonumber \\
S/\sqrt{S+B} \ &= \ 1.6 \ \ \ {\rm for} \ (m_{\tilde{t}_1}, \, m_{\tilde{\chi}_1^0})=(323,~148)~{\rm GeV},
 \ {\rm with \ } 300~{\rm fb}^{-1} \ {\rm of \ data.} 
\end{align}
We can further improve the significance by taking advantage of the peak distribution of signal events
 on ($\Delta \phi( \vec{\mpt}_T, \, \vec{p}_{t\bar{t} \, T} ), \, \mpt_T/p_{t\bar{t} \, T}$) plane,
 but this is beyond the scope of our theoretical study.
\\

We have studied the equal-velocity scenario,
 in which a \textit{stop} and a neutralino are the lightest SUSY particles
 and the \textit{stop} mass is close to the sum of the SM top quark mass and 
 the neutralino mass, which forces
 the top quark and the neutralino coming from the decay of a \textit{stop}
 to have similar velocity vectors.
In this case, it is difficult to distinguish 
 \textit{stop} pair production events from the SM top quark pair production background
 because missing transverse momentum due to neutralinos is suppressed.
We have proposed a novel analysis technique aiming at discovering the \textit{stop} signal
 in the equal-velocity scenario,
 which takes advantage of the information that, in the equal-velocity limit,
 missing transverse momentum vector due to the neutralinos
 and the transverse momentum vector of the SM top quark pair
 have the same direction
 and the ratio of their absolute values equals to the ratio of the neutralino mass
 and the top quark mass.
We have investigated the distributions of \textit{stop} pair production events
 and SM top quark pair production events that pass special selection cuts on the reconstructed top quarks,
 on the plane of
 $( \Delta \phi(\vec{p}_{t\bar{t} \, T}, \vec{\mpt}_T), \ \mpt_T / p_{t\bar{t} \, T} )$.
We have found that even after a full detector simulation, \textit{stop} events gather around the point of 
 $\Delta \phi(\vec{p}_{t\bar{t} \, T}, \vec{\mpt}_T)=0$ and $\mpt_T / p_{t\bar{t} \, T} = m_{\tilde{\chi}_1^0}/m_t$,
 while the SM top quark background does not show such a tendency.
Hence we expect that the event scatter plot on the plane of
 $( \Delta \phi(\vec{p}_{t\bar{t} \, T}, \vec{\mpt}_T), \ \mpt_T / p_{t\bar{t} \, T} )$
 leads to the discovery of a \textit{stop} signal if the mass spectrum is in accord with the equal-velocity scenario.
\\

\end{document}